%% file: main.tex
\documentclass[superscriptaddress,preprintnumbers,amsmath,amssymb,twocolumn,floats,aps,prl]{revtex4-1}
\usepackage{txfonts}
\usepackage{amssymb}
\usepackage{graphicx}
\usepackage{epstopdf}
\usepackage{sidecap}
\usepackage{CJK}
\usepackage{txfonts}
\usepackage{url}

\begin{document}

\bibliographystyle{unsrt}
\title{Spontaneous ferromagnetism induced topological phase transition in EuB${_6}$}

\author{W. L. Liu}
\thanks{Equal contributions}
\affiliation{Center for Excellence in Superconducting Electronics, State Key Laboratory of Functional Materials for Informatics, Shanghai Institute of Microsystem and Information Technology, Chinese Academy of Sciences, Shanghai 200050, China}
\affiliation{Center of Materials Science and Optoelectronics Engineering, University of Chinese Academy of Sciences, Beijing 100049, China}
\affiliation{School of Physical Science and Technology, ShanghaiTech University, Shanghai 201210, China}

\author{X. Zhang}
\thanks{Equal contributions}
\affiliation{School of Physical Science and Technology, ShanghaiTech University, Shanghai 201210, China}

\author{S. M. Nie}
\thanks{Equal contributions}
\affiliation{Department of Materials Science and Engineering, Stanford University, Stanford, California 94305, USA}

\author{Z. T. Liu}
\thanks{Equal contributions}
\affiliation{Center for Excellence in Superconducting Electronics, State Key Laboratory of Functional Materials for Informatics, Shanghai Institute of Microsystem and Information Technology, Chinese Academy of Sciences, Shanghai 200050, China}

\author{X. Y. Sun}
\affiliation{School of Physical Science and Technology, ShanghaiTech University, Shanghai 201210, China}

\author{H. Y. Wang}
\affiliation{School of Physical Science and Technology, ShanghaiTech University, Shanghai 201210, China}

\author{J. Y. Ding}
\affiliation{Center for Excellence in Superconducting Electronics, State Key Laboratory of Functional Materials for Informatics, Shanghai Institute of Microsystem and Information Technology, Chinese Academy of Sciences, Shanghai 200050, China}
\affiliation{Center of Materials Science and Optoelectronics Engineering, University of Chinese Academy of Sciences, Beijing 100049, China}

\author{L. Sun}
\affiliation{School of Information Science and Technology, ShanghaiTech University, Shanghai 201210, China}

\author{Z. Huang}
\affiliation{Center for Excellence in Superconducting Electronics, State Key Laboratory of Functional Materials for Informatics, Shanghai Institute of Microsystem and Information Technology, Chinese Academy of Sciences, Shanghai 200050, China}

\author{H. Su}
\affiliation{School of Physical Science and Technology, ShanghaiTech University, Shanghai 201210, China}

\author{Y. C. Yang}
\affiliation{Center for Excellence in Superconducting Electronics, State Key Laboratory of Functional Materials for Informatics, Shanghai Institute of Microsystem and Information Technology, Chinese Academy of Sciences, Shanghai 200050, China}

\author{Z. C. Jiang}
\affiliation{Center for Excellence in Superconducting Electronics, State Key Laboratory of Functional Materials for Informatics, Shanghai Institute of Microsystem and Information Technology, Chinese Academy of Sciences, Shanghai 200050, China}

\author{X. L. Lu}
\affiliation{Center for Excellence in Superconducting Electronics, State Key Laboratory of Functional Materials for Informatics, Shanghai Institute of Microsystem and Information Technology, Chinese Academy of Sciences, Shanghai 200050, China}
\affiliation{Center of Materials Science and Optoelectronics Engineering, University of Chinese Academy of Sciences, Beijing 100049, China}

\author{J. S. Liu}
\affiliation{Center for Excellence in Superconducting Electronics, State Key Laboratory of Functional Materials for Informatics, Shanghai Institute of Microsystem and Information Technology, Chinese Academy of Sciences, Shanghai 200050, China}
\affiliation{Center of Materials Science and Optoelectronics Engineering, University of Chinese Academy of Sciences, Beijing 100049, China}

\author{Z. H. Liu}
\affiliation{Center for Excellence in Superconducting Electronics, State Key Laboratory of Functional Materials for Informatics, Shanghai Institute of Microsystem and Information Technology, Chinese Academy of Sciences, Shanghai 200050, China}
\affiliation{Center of Materials Science and Optoelectronics Engineering, University of Chinese Academy of Sciences, Beijing 100049, China}

\author{S. L. Zhang}
\affiliation{School of Physical Science and Technology, ShanghaiTech University, Shanghai 201210, China}

\author{H. M. Weng}
\affiliation{Institute of Physics and Beijing National Laboratory for Condensed Matter Physics, Chinese Academy of Sciences, Beijing 100190, China}
\affiliation{CAS Center for Excellence in Topological Quantum Computation, University of Chinese Academy of Sciences, Beijing 100049, China}

\author{Y. F. Guo}
\email{guoyf@shanghaitech.edu.cn}
\affiliation{School of Physical Science and Technology, ShanghaiTech University, Shanghai 201210, China}

\author{Z. J. Wang}
\email{zjwang@iphy.ac.cn}
\affiliation{Institute of Physics and Beijing National Laboratory for Condensed Matter Physics, Chinese Academy of Sciences, Beijing 100190, China}
\affiliation{CAS Center for Excellence in Topological Quantum Computation, University of Chinese Academy of Sciences, Beijing 100049, China}

\author{D. W. Shen}
\email{dwshen@mail.sim.ac.cn}
\affiliation{Center for Excellence in Superconducting Electronics, State Key Laboratory of Functional Materials for Informatics, Shanghai Institute of Microsystem and Information Technology, Chinese Academy of Sciences, Shanghai 200050, China}
\affiliation{Center of Materials Science and Optoelectronics Engineering, University of Chinese Academy of Sciences, Beijing 100049, China}

\author{Z. Liu}
\affiliation{School of Physical Science and Technology, ShanghaiTech University, Shanghai 201210, China}

\begin{abstract}
The interplay between various symmetries and electronic bands topology is one of the core issues for topological quantum materials. Spontaneous magnetism, which leads to the breaking of time-reversal symmetry, has been proven to be a powerful approach to trigger various exotic topological phases. In this work, utilizing the combination of angle-resolved photoemission spectroscopy, magneto-optical Kerr effect microscopy and first-principles calculations, we present the direct evidence on the realization of the long-sought spontaneous time-reversal symmetry breaking induced topological phase transition in soft ferromagnetic EuB$_6$. We successfully disentangle the bulk band structure from complicated surface states, and reveal the hallmark band inversion occurring between two opposite-parity bulk bands below the phase transition temperature. Besides, our magneto-optical Kerr effect microscopy result confirms the simultaneous formation of magnetic domains in EuB$_6$, implying the intimate link between the topological phase transition and broken time-reversal symmetry therein. Our results demonstrate that EuB$_6$ provides a potential platform to study the interplay between the topological phases and tunable magnetic orders.
\end{abstract}

\maketitle

In the past decades, the investigation of topological phases has aroused a great surge of interest due to both the breakthrough on the paradigm of condensed matter physics research and various potential applications. One core issue on these research activities is to understand the interplay between all kinds of symmetries and topology~\cite{RN240}. In point of fact, topological non-trivial materials are defined as a specific class of materials in which electronic structure can be classified by topological invariants protected by various symmetries~\cite{RN252,RN242,RN128,RN38}. Among them, time-reversal symmetry (TRS) is the one which received the first and also most attention. In topological insulators (TIs), it is the TRS that protects novel gapless helical surface states~\cite{RN41,2009Experimental,RN786,RN367}. Besides, the breaking of TRS can significantly alter the electronic structure and then give rise to exotic topological phase transitions, such as the formation of magnetic Weyl nodes from topological trivial bands through the application of external magnetic field in half-Heusler compounds~\cite{RN676,RN787,RN788}. Particularly, the searching for topological phase transitions due to spontaneous broken TRS is of particular interest taking into account that the interaction between intrinsic magnetism and topology may introduce a variety of novel physics~\cite{0Intrinsic,2020Realization,RN791,RN802,2101.10149,nie2017topological,hua2018dirac}. For example, two-dimensional (2D) ferromagnetic or antiferromagnetic TIs, in which the TRS breaking can lead to the topological phase transition to quantum anomalous Hall insulators (QAHI)~\cite{2010Quantized,2013Experimental,RN800,RN789,nie2019topological} or topological axion insulators~\cite{RN530,RN803,RN801,RN432,RN790,RN783}, have been extensively studied. In sharp contrast, topological phase transitions directly from topological trivial to non-trivial phase due to spontaneous broken TRS are rare, and it is highly desired to visualize this phase transition.

Recently, it was suggested that europium hexaboride (EuB$_6$), a typical soft magnetic material, should undergo a topological phase transition from a small gap semiconductor in its paramagnetic (PM) state to a TRS broken topological semimetal in the ferromagnetic (FM) state, considering that the effective magnetic exchange splitting renders opposite effect on the two bands and consequently leads to the band inversion in the spin-up subbands~\cite{RN212}. This material provides an ideal platform to realize the long-sought topological phase transition caused by spontaneous TRS breaking.  Previous angle-resolved photoemission spectroscopy (ARPES) studies on EuB$_6$ just suggested an $X$-point band gap of as large as 1 eV with the Fermi level ($E_F$) near the bottom of the conduction band~\cite{RN805,2001Absence} and no sign of the band inversion, in conflict with the semimetal character with a small $X$-point band overlap revealed by bulk-sensitive techniques, such as the quantum oscillation~\cite{1998Fermi,M1999Fermi}. Besides, EuB$_6$ has drawn much attention due to its interesting transport properties around magnetic transition temperatures as well~\cite{B1998Electron,1997Low,RN796}. It becomes ferromagnetic below $T_{c2}$ =12.5 K, with the moment oriented to the [111] direction. A large negative magnetoresistance, which is related to percolation-type transition resulting from the overlap of magnetic polarons, occurs at another magnetic phase transition temperature $T_{c1}$ =15.3 K~\cite{1998Structure,RN798,RN797,S2000Metallization,RN799,2004Magnetic,1980Specific}. Moreover, below $T_{c1}$, a mysterious metal-insulator transition with one order magnitude drop of the resistivity in zero field takes place, and so far the underlying mechanism is still unclear~\cite{RN776,1997Spectroscopic}. The experimental exploration of the low-lying electronic structure of EuB$_6$ would shed light on the understanding of all above mysterious topological and electrical transport properties therein.


In this Letter, combining angle-resolved photoemission spectroscopy (ARPES), magneto-optical Kerr effect (MOKE) microscopy and first-principles calculations, we systematically investigated the evolution of electronic structure and magnetic domains of EuB$_6$ when it undergoes the FM phase transition. We have provided a direct evidence for the realization of topological phase transition in EuB$_6$ via successfully disentangling its three-dimensional bulk band structure from complicated surface states for the first time, which unveils that the band inversion occurs between two opposite-parity bulk bands in the spin-up channel, together with the coincident appearance of magnetic domains. Moreover, the long-standing puzzle of metal-insulator transition in EuB$_6$ could also be interpreted by our findings.


\begin{figure}[t]
\includegraphics[width=8.3cm]{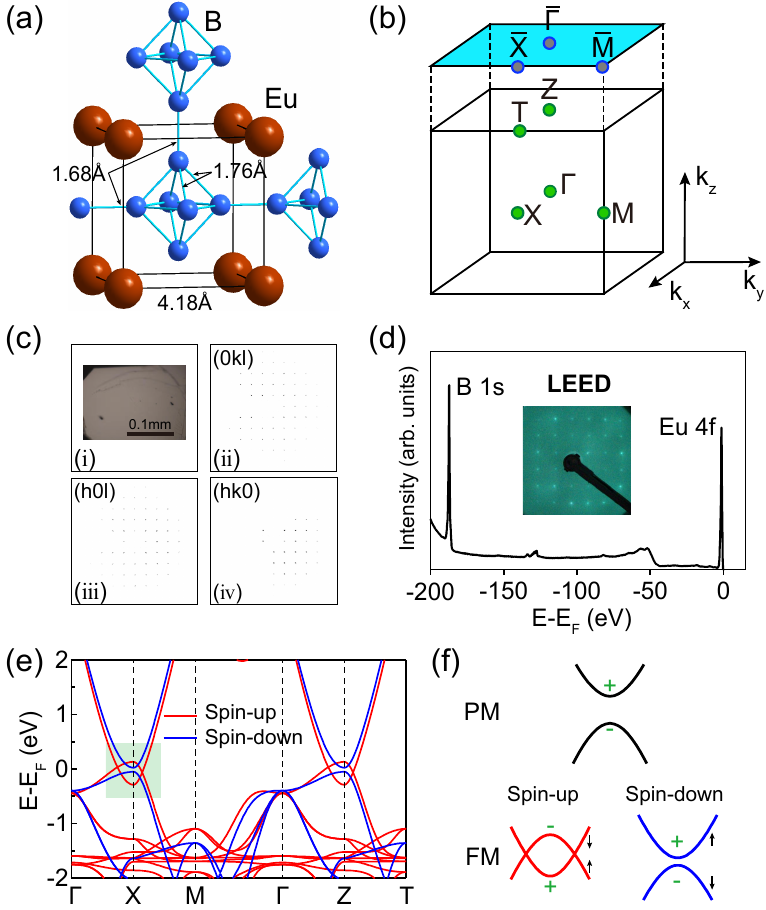}
\caption{(a) The crystal structure of EuB$_6$. (b) Bulk and (001)-projected surface Brillouin zones of EuB$_6$. (c) ($\romannumeral1$) The image of high quality cleaved EuB$_6$ single crystal. ($\romannumeral2$)-($\romannumeral4$) are X-ray diffraction patterns from the ($0kl$), ($h0l$) and ($hk0$) surfaces, respectively. (d) The core-level photoemission spectrum of EuB$_6$ single crystal, showing characteristic Eu and B energy levels. The inset presents the LEED pattern of EuB$_6$ (001) cleavage surface. (e) The calculated band structure of EuB$_6$ in its ferromagnetic state. (f) The schematic plot of the band inversion occurring at the $X$ point.}
\label{crystal structure}
\end{figure}

High-quality single crystals of EuB$_6$ were synthesized by using aluminum as the flux. Starting materials Eu rods (99.9$\%$, Alfa Aesar), B powder (99.9$\%$, Macklin), and Al rods (99.9$\%$, Macklin) were mixed in the molar ratio 1 : 6 : 120 and sealed into an alumina crucible. All the procedures were performed in a glove box protected with high purity argon gas. Then the crucible was slowly heated up to 1623 K in a high temperature vacuum atmosphere furnace, kept at the temperature for 20 hours, and followed by cooling down to 923 K at a temperature decreasing rate of 6 K/h. After taken out of the crucible, crystals were separated from the flux by leaching in HCl. The phase purity and quality of the crystals were examined on a single crystal X-ray diffractometer equipped with a Mo K$\alpha$ radioactive source ($\lambda$ = 0.71073~$\AA$). 

High-resolution ARPES measurements were performed at 03U and ``Dreamline" beam lines of Shanghai Synchrotron Radiation Facility (SSRF)~\cite{2102.09915}. These two endstations are equipped with Scienta-Omicron DA30 and DA30-L electron analyzers, respectively. The angular and the energy resolutions were set to 0.2$^\circ$ and 8$\sim$20 meV (dependent on the selected probing photon energy). All samples were cleaved in an ultrahigh vacuum better than 8.0 $\times$ 10$^{-11}$ Torr.


The first-principles calculations were performed with the projector augmented wave method implemented in Vienna ab initio simulation package~\cite{kresse1996efficient,kresse1996efficiency}. Exchange and correlation potential was treated within the generalized gradient approximation (GGA) of Perdew–Burke–Ernzerhof type~\cite{perdew1996generalized}. The cut-off energy for plane wave expansion was 500 eV. The k-point sampling grid in the self-consistent process was set to 6 $\times$ 6 $\times$ 1. Spin orbital coupling was included in the calculations. The GGA+Hubbard-U method~\cite{liechtenstein1995density} was used to treat the correlation effect in EuB$_6$ with U = 7 eV for the $f$-electrons of Eu.


EuB$_6$ has a CsCl-type structure with space group Pm3m (No. 221) as shown in Fig.~\ref{crystal structure}(a)~\cite{2001Temperature}. There exist two possible cleavage surfaces parallel to the (001) plane, one between adjacent Eu/B planes and the other between two B planes within an octahedron, resulting in either polar or electrically neutral terminations~\cite{2020Visualization}. During our ARPES measurements on more than 50 samples, we have come across both cases. In addition, the bulk and (001)-projected surface Brillouin zones (BZs) of EuB$_6$ are shown in Fig.~\ref{crystal structure}(b). Our samples were characterized by the single-crystal X-ray diffraction [Fig.~\ref{crystal structure}(c)], and the results along high-symmetry directions show individual dots with no impurity phases appearing, suggesting the excellent quality of the sample. Fig.~\ref{crystal structure}(d) displays the angle-integrated photoemission spectrum of EuB$_6$ over a large range of binding energy, in which we can clearly identify the Eu (\emph{4f}) and B (\emph{1s}) energy levels, confirming the element composition of our samples. After cleaved in the air, the sample shows typical flat and shining surface as illustrated in the inset of Fig.~\ref{crystal structure}(c). Meanwhile, the low-energy electron diffraction (LEED) pattern, shown in the inset of Fig.~\ref{crystal structure}(d), confirms the square (001) cleavage surface of EuB$_6$.


Our first-principles calculation shows that PM EuB$_6$ is a semiconductor. After the consideration of FM order, the spin-up occupied $f$ states of Eu are located around -1.5 eV and the spin-down unoccupied $f$ states of Eu are high above $E_F$, as shown in Fig.~\ref{crystal structure}(e). In the spin-up (spin-down) channel, the highest valence band dominated by $p$ states of B is pushed upward (downward) due to the hybridization with the occupied (unoccupied) $f$ states, leading to an effective antiferromagnetic exchange coupling in the highest valence band, and the band inversion at three $k$-points (including $X$, $Y$ and $Z$ points, which are equivalent under $\hat{c}_{3}^{111}$ symmetry). For the magnetic centrosymmetric system, the topological invariant $\chi$ is defined by
\begin{equation}
(-1)^\chi\equiv\prod_{j=\{1,2,\cdots,n_{occ}\},~\Gamma_i=\text{TRIMs}} \xi^j_{i}
\end{equation}
where $\xi_i^j$ is the parity eigenvalue of the $j$th band at the time-reversal-invariant-momentum (TRIM) $\Gamma_i$. If $\chi=1$, the system cannot be fully gapped and must have band crossing points around $E_F$, forming a typical topological semimetal. Generally, it can be either a nodal-line semimetal or a Weyl semimetal according to its magnetic symmetries. When the magnetic moment is aligned along the [001] direction, EuB$_6$ is a nodal line semimetal with three nodal lines circled around $X$, $Y$ and $Z$, respectively. The nodal lines are protected by the mirror symmetry $M_z$. 


\begin{figure}[t]
\includegraphics[width=8.3cm]{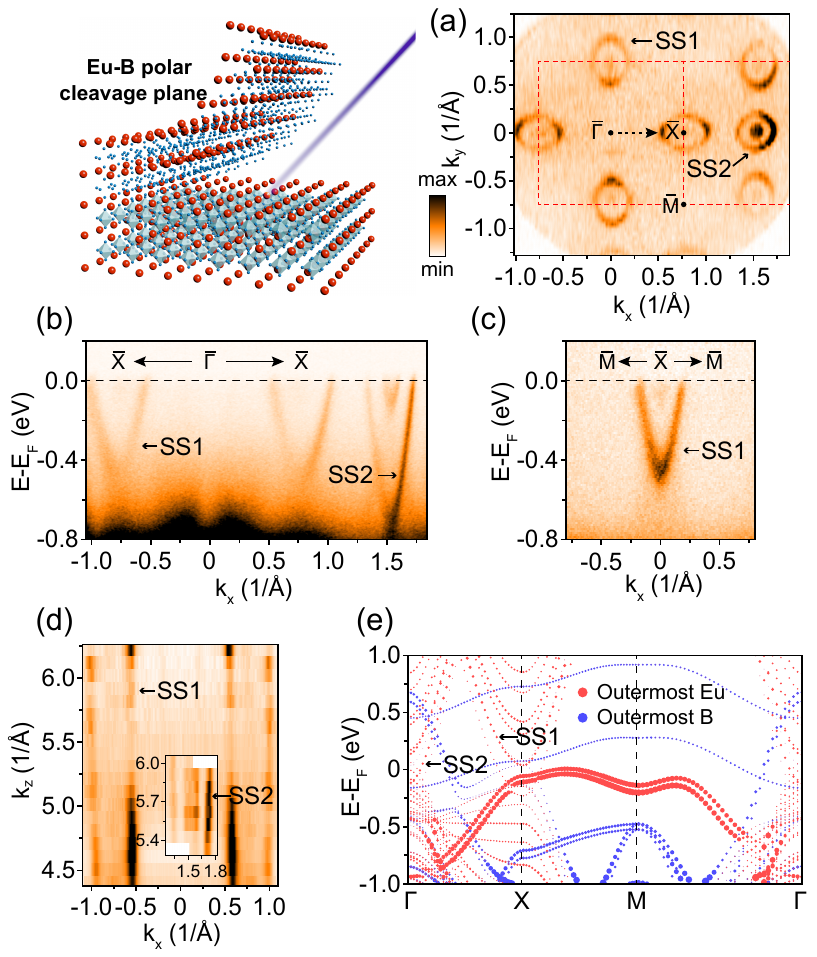}
\caption{(a) Photoemission intensity map for Eu-B polar cleavage plane, recorded with linearly horizontally polarized 116 eV photons at 9.5 K. The intensity was integrated at $E_F$ with an energy window of 50 meV. $\overline{\Gamma}$, $\overline{X}$ and $\overline{M}$ represent high-symmetry points in the projected two-dimensional BZ. (b) and (c) Photoemission intensity plots along $\overline{\Gamma}$-$\overline{X}$ and $\overline{M}$-$\overline{X}$, respectively. (d) 
Photoemission intensity map on the \emph{k$_x$-k$_z$} plane taken at \emph{$E_F$}. Inset is the intensity map for the band SS2. (e) First-principles calculations for a nine-layered slab. The sizes of the red and blue circles denote the weight projection of the outermost Eu layer and B layer, respectively.}
\label{ARPES_surface}   
\end{figure}

\begin{figure*}[htbp]
\centering
\includegraphics[width=18cm]{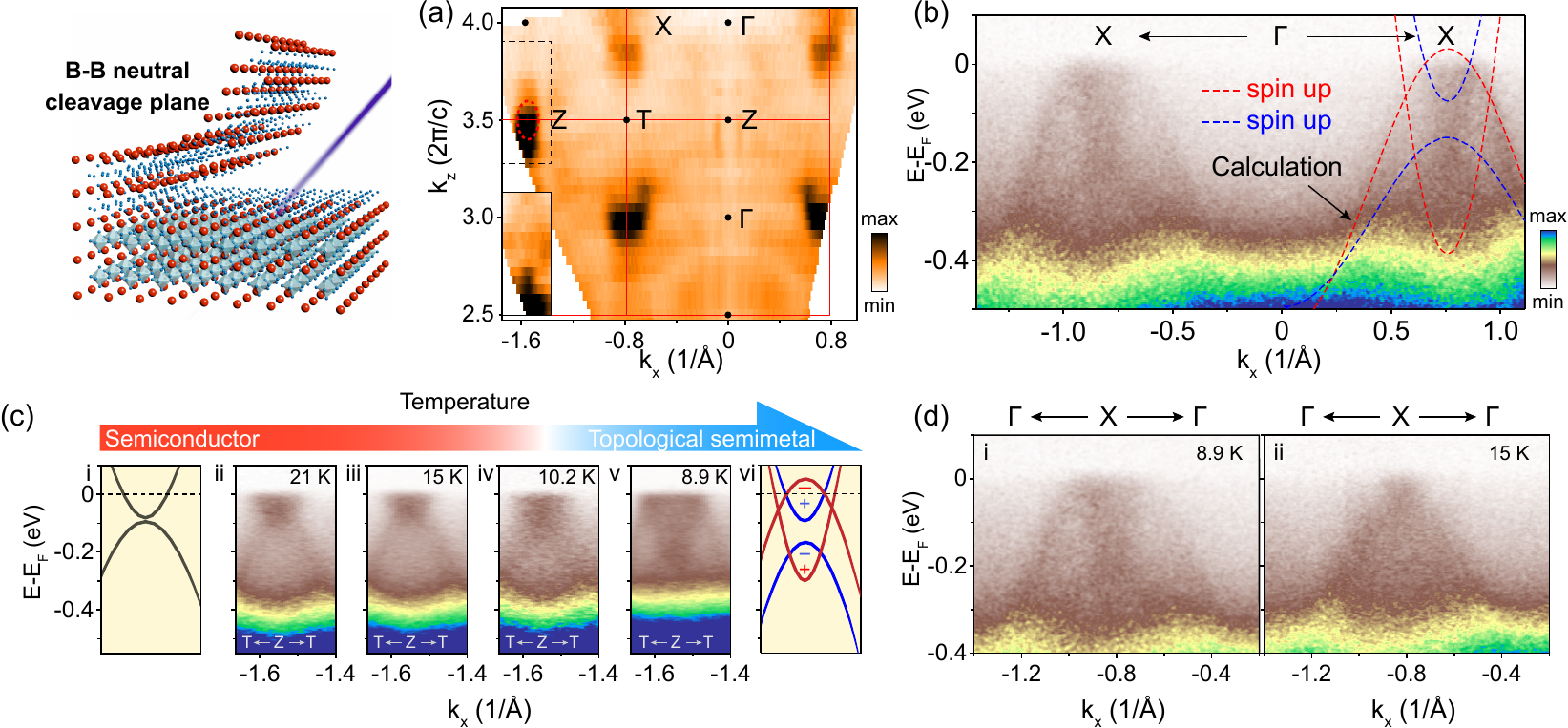}
\caption{(a) Photoemission intensity maps on the \emph{k$_x$-k$_z$} plane taken around \emph{$E_F$} at 8.9 K for the neutral B-B cleavage plane. The photon energy is ranged from 40 to 136 eV. The inner potential of 15 eV was determined to match the periodicity along $k_Z$. Inset is the intensity map taken at \emph{$E_F$} - 0.2 eV for the dashed square containing the Z point, suggesting the 3D character of the valence band. (b) Photoemission intensity plot along $\varGamma$-$X$ taken with 130 eV photons. Dished lines represent the band structure calculation for comparison. (c) $\romannumeral2$-$\romannumeral5$ record the developing of band inversion around the $X(Z)$ point along $T$-$Z$-$T$ with the temperature. $\romannumeral1$ and $\romannumeral6$ show schematic plots of band structure above and below $T_{c}$, respectively. (d) Photoemission intensity plots taken along $\varGamma$-$X$ at 8.9~K and 15~K.}
\label{ARPES_bulk}
\end{figure*}

Previous ARPES studies on EuB$_6$ have revealed unstable surface states with prominent temporal change of chemical potential~\cite{2001Absence}. Such complicated surface states would inevitably hide the photoemission signal from real bulk states. We performed comprehensive survey on the three-dimensional electronic structure of EuB$_6$ through synchrotron radiation based mirco-ARPES (with a tiny $30~\times~30$ $\mu$m beam spot) and numerous attempts of cleavage, and eventually we achieved two sets of data taken on different cleavage planes: one from the polar Eu/B termination and the other from the electrical neutral B/B termination, which is the key to the success in digging out the unreported bulk band structure of EuB$_6$.

Figure~\ref{ARPES_surface} shows the photoemission results taken from the polar Eu/B termination. We discovered the similar temporal change of chemical potential to previous ARPES reports, as illustrated in Figs. S1 and S2, and Note 1 of Supplemental Material (SM)~\cite{si}. We can assign this phenomenon to the atomic surface absorption effect. In Figs.~\ref{ARPES_surface}(a-c), we present the surface map and photoemission intensity plots along high-symmetry directions taken in the FM state (9.5~K) after the adsorption saturation, in which two concentric circular electron pockets and two elliptical ones appear around the BZ center and boundary, respectively. Here, due to the matrix element effect, the circular pockets are more clearly visible in the second BZ. Figure~\ref{ARPES_surface}(d) and its inset present intensity maps on the \emph{k$_x$-k$_z$} plane taken at \emph{$E_F$} for the elliptical and circular pockets, respectively. Our results demonstrate negligible \emph{k$_z$} dispersions for all these bands, suggesting their surface-state character. According to our slab calculation [Fig.~\ref{ARPES_surface}(e)], we can attribute them mainly to the electron bands originating from Eu $5d$ orbitals above $E_F$ (considering the upward shift of chemical potential after the adsorption saturation). Here, we note that there exists exchange splitting for these surface bands.

According to theoretical prediction, when undergoing to the FM state, the aligned magnetic moments in EuB$_6$ have opposite effective exchange splitting on the low-energy bulk bands. Thus, the band inversion would occur only in the spin-up channel between two opposite-parity bulk bands at $X(Y, Z)$ points, leading to the non-trivial topological phase transition from a normal semiconductor to a topological semimetal, as schematically illustrated in Fig.~\ref{crystal structure}(f). The key to confirming this prediction is to straightforwardly observe the splitting and crossing evolution of these bulk bands upon temperature.

We successfully obtained the electrically neutral B/B cleavage planes and thus collected the more stable photoemission spectra [Fig.~\ref{ARPES_bulk}], in which we did not find any complicated temporal changes. Figure~\ref{ARPES_bulk}(a) and its inset illustrate photoemission intensity maps on the \emph{k$_x$-k$_z$} plane taken around \emph{$E_F$} and \emph{$E_F$} - 0.2 eV in the FM state (8.9~K) for this neutral surface, respectively. Both data present clear dispersions for bands around $X$($Z$) points with periodic modulation along \emph{$k_z$}, confirming their bulk-band nature. The photoemission intensity plot along the high-symmetry direction $X$-$\varGamma$-$X$ is demonstrated in Fig.~\ref{ARPES_bulk}(b), through which crossings between two splitted hole-like valence bands and the lower conduction band can be well resolved. Although the exchange splitting of original conduction band in the PM state can not be distinguished due to the intrinsic photoemission $k_z$ broadening effect for bulk bands, the predicted band inversion between two opposite-parity bands resulting in $\chi=1$ can be confirmed here. Furthermore, we performed temperature dependent ARPES measurements on these bulk bands to investigate the detailed evolution of topological phase transition. Figures~\ref{ARPES_bulk}(c)($\romannumeral2$-$\romannumeral5$) illustrate the developing of band inversion around the $X(Z)$ point along $T$-$Z$-$T$ with the temperature. Above 15~K, the conduction and valence bands barely touch with each other, forming a typical normal semiconductor. While, just below a critical temperature around 10.2~K, the band exchange splitting begins springing up and the splitted lower conduction band (spin-up channel subband) starts to sink and gradually cross with both splitted valence bands. This phenomenon is more apparent for the spectral comparison taken along $\varGamma$-$X$-$\varGamma$~[Figs.~\ref{ARPES_bulk}(d)($\romannumeral1$-$\romannumeral2$)], in which the magnitude of such band splitting is larger due to the band anisotropy. Such band structure evolution is in remarkable agreement with the theoretical prediction [Fig.~\ref{crystal structure}(f) and Fig.~\ref{ARPES_bulk}(c)], and thus the topological phase transition in EuB$_6$ can be experimentally confirmed. Moreover, the evolution of bulk band crossings occurring in EuB$_6$ as well naturally explains the long-standing puzzle of metal-insulator transition therein.

\begin{figure}[t]
\includegraphics[width=8.3cm]{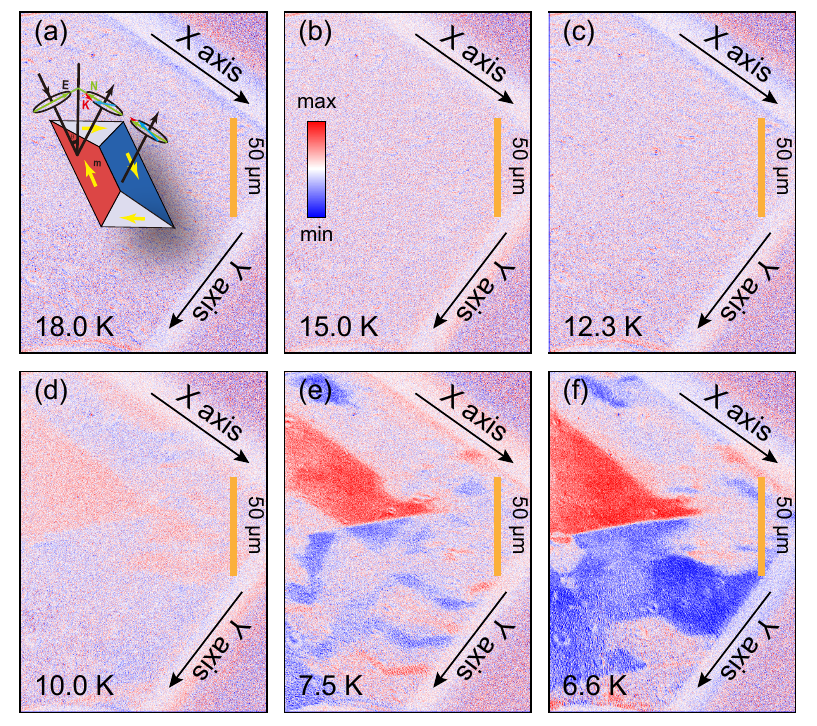}
\caption{(a)-(f) MOKE microscopy images at various temperature recorded on pure long mode (Horizontal direction in the figure). $X$ and $Y$ axis indicate the nature edge of the sample. Inset in (a2) is the schematic of magneto-optic kerr effect. The shape of sample measured here is a cubic with the size about 120 $\mu$m $\times$ 120 $\mu$m $\times$ 120 $\mu$m.}
\label{MOKE}
\end{figure}

Next, to confirm the bulk band evolution indeed induced by spontaneous magnetization, and to research the magnetism behavior of EuB$_6$, we applied the MOKE microscopy as a non-destructive probe for imaging the evolution of magnetic domains with temperature and their corresponding dynamics, as shown in Fig.~\ref{MOKE}. Figures~\ref{MOKE}(a)-(f) present a series of optical intensity difference plot in the zero field on the cleaved EuB${_6}$ (001) plane (see details in Fig. S3 and Note 2 of SM)~\cite{si}. Above 12.3 K, no net magnetic moment signal can be resolved. Below 10 K, however, as the temperature goes down, the MOKE microscopy image presents a more and more striking contrast (red versus blue) between different areas in the sample region, suggesting the formation of individual magnetic domains and thus the breaking of TRS. We note that such domains begin springing up during 10.0~$\sim$~12.3~K, in remarkably consistent with the temperature at which the bulk band splitting starts to occur [Fig.~\ref{ARPES_bulk}(c)]. This means that the band inversion between two opposite-parity bulk bands in EuB${_6}$ is closely relevant to the spontaneous TRS breaking therein.

In summary, we have systematically investigated the evolution of electronic structure and magnetic domains of EuB$_6$ when it undergoes the FM transition. We successfully disentangle its three-dimensional bulk band structure from complicated surface states for the first time. The revealed band inversion occurring in the spin-up channel of bulk bands, together with the coincident appearing of magnetic domains, provides a direct evidence on the realization of topological phase transition in EuB$_6$.

\begin{acknowledgments}
This work was supported by the National Key R$\&$D Program of the MOST of China (Grant No. 2016YFA0300204), the National Science Foundation of China (Grant Nos. U2032208, 11874264), and the Natural Science Foundation of Shanghai (Grant No. 14ZR1447600). Y. F. Guo acknowledges the starting grant of ShanghaiTech University and the Program for Professor of Special Appointment (Shanghai Eastern Scholar). Part of this research used Beamline 03U of the Shanghai Synchrotron Radiation Facility, which is supported by ME$^2$ project under contract No. 11227902 from National Natural Science Foundation of China. The authors also thank the support from Analytical Instrumentation Center (\#SPST-AIC10112914), SPST, ShanghaiTech University. 
\end{acknowledgments}


\input{main.bbl}

\bibliographystyle{apsrev4-1}

\end{document}

%% file: main.bbl
%